\documentclass[a4paper,10pt]{article}
\usepackage{graphicx}
\usepackage{jcappub}

\title{Testing the Dipole Modulation Model in CMBR}

\author[a]{Pranati K. Rath}
\author[a]{Pankaj Jain}

\affiliation[a]{Dept. of Physics, Indian Institue of Technology Kanpur, Kanpur - 208016, India}

\emailAdd{pranati@iitk.ac.in}
\emailAdd{pkjain@iitk.ac.in}

\abstract{
The hemispherical power asymmetry, observed in the CMBR data, has
generally been interpreted in terms of the dipole modulation model
for the temperature fluctuations. Here we point out that this model
leads to several predictions, which can be directly tested in the
current data. We suggest tests of the hemispherical power asymmetry
both in real and multipole space. We find a significant signal of the
dipole modulation model in WMAP and PLANCK data with our tests.
The dipole amplitude and direction also agrees, within errors, with earlier
results based on hemispherical analysis in multipole space.
We also find evidence that the effective dipole modulation amplitude increases
with the multipole $l$ in the range $l=2-64$. 
}

\arxivnumber{1308.0924}

\begin{document}

\maketitle

\section{Introduction}

At present there exists considerable evidence which indicates that
the Universe may not be isotropic on large distance scales. 
For example, 
there are several observations which suggest a preferred axis pointing
roughly towards Virgo \cite{Jain1999,Hutsemekers1998,Costa2004,Ralston2004,Schwarz2004,Singal2011}. One also observes a hemispherical anisotropy \cite{Eriksen2004,Eriksen2007,Hansen2009,Erickce2008b,Paci2013,Schmidt2013}, where the power extracted from two different hemispheres shows
significant difference from one another.
 The power in each hemisphere is estimated by making a harmonic decomposition
of the masked sky. The significance of anisotropy is found to be about 3 sigma \cite{planck2013}
with the axis of maximum asymmetry having $(\theta,\phi)$ pointing towards $(115^{\circ}, 227^\circ)$ in galactic coordinates. 
This direction is nearly perpendicular to the direction towards Virgo. 
These observations suggest a violation of the cosmological principle
and might indicate that one needs a revision of the Big Bang cosmology.
There exist many models which propose to address these issues.
It has been suggested that the anisotropic modes, generated during the
pre-inflationary phase might re-enter the horizon before the current time
\cite{Aluri2012,Pranati2013,Chang2013}. Hence these might explain the observed 
anisotropy even within the framework of the inflationary Big Bang model.

Theoretically, the observed hemispherical asymmetry in CMBR is 
assumed to arise from the model \cite{Gordon2005,Gordon2007,Prunet2005,Bennett2011},  
\begin{equation}
 {\bigtriangleup T}(\hat n) = g(\hat n) \left(1+A \hat \lambda \cdot \hat n \right)
\label{eq:dipole_mod}
\end{equation}
where ${\bigtriangleup T}(\hat n)$ is the measured temperature 
fluctuation in  the direction $\hat n$, $ g(\hat n)$ a statistically
isotropic field, $A$ the dipole amplitude and 
$\hat \lambda$ the preferred direction. 
Taking the preferred direction along the z-direction, the temperature modulated field can be written as,
\begin{equation}
 {\bigtriangleup T}(\hat n) = g(\hat n) 
\left(1+A \cos\theta\right)
\end{equation}
The extracted value of the dipole modulation amplitude from 
CMBR data is given by 
\cite{Eriksen2004,Eriksen2007,Hansen2009,Erickce2008b,Paci2013,Schmidt2013}, 
$A=0.072\pm 0.022$. 
The precise value of $A$ shows some dependence on the input CMBR map. 
However the difference is relatively small and not relevant
for our analysis.
It is likely that this model may not extend to all values of $l$. 
The true anisotropy model is likely to be more complicated \cite{Liu2013,Cai2013,Flender2013}.
Here we discuss several implications of this model and test them in the 
observed CMBR data. These provide additional tests of this model.

\section{Tests of the Dipole Modulation Model}
In this section we suggest several tests of the dipole modulation model,
both in pixel and multipole space. 
The temperature field, $\Delta T(\hat n)$ can be analyzed by making a spherical
harmonic decomposition,
\begin{equation}
 {\bigtriangleup T}(\hat n) = \sum_{l = 1}^{\infty}\sum_{m = -l}^{+l}a_{lm}Y_l^m(\hat n) 
\label{eq:DeltaT_exp}
\end{equation}
Similarly, 
we expand the statistically isotropic field, $ g(\hat n)$, 
\begin{equation}
 g(\hat n) = \sum_{l,m }\tilde a_{lm}Y_l^m(\hat n) 
\label{eq:gn_exp}
\end{equation}
The expansion coefficients, $\tilde a_{lm}$, in this case are different 
from $a_{lm}$ due to the presence of dipole modulation in $\Delta T$.  
However, the multipole power, $C_l$, is same for both the fields, ${\bigtriangleup T}(\hat n)$ and $g(\hat n)$. We shall show this explicitly in Section 2.2. 
We next describe our tests of the dipole modulation in real space, followed
by tests in multipole space.

\subsection{Tests in real space}
Let us consider the observed squared temperature fluctuation field, 
\begin{equation}
f(\theta,\phi) = 
\left({\Delta T(\theta,\phi)}\right)^2
\end{equation}
where we have expressed the unit vector, $\hat n$, in terms of the
spherical polar coordinates, $(\theta,\phi)$.
If the dipole modulation model, Eq. \ref{eq:dipole_mod}, provides an
accurate description of the real data, we expect that $f(\theta, \phi)$
can be expressed as,
\begin{equation}
f(\theta,\phi)
\approx 
g(\theta,\phi) g^*(\theta,\phi)
\left(1+2 A\hat\lambda\cdot\hat n\right)
\end{equation}
where on the right hand side we have only kept terms linear in the dipole
amplitude $A$. 
Using Eq. \ref{eq:gn_exp}
 and taking the ensemble average, we obtain, 
\begin{equation}
\langle f(\theta,\phi)\rangle
\approx 
\sum_{l,m}\sum_{l'm'} \langle \tilde a_{lm} \tilde a^*_{l'm'} \rangle Y_l^m(\theta,\phi)
Y^{m'*}_{l'}(\theta,\phi) 
\left(1+2 A\hat\lambda\cdot\hat n\right)
\end{equation}
Using 
\begin{equation}
\langle \tilde a_{lm} \tilde a^*_{l'm'} \rangle = \delta_{ll'}\delta_{mm'} C_l
\end{equation}
and the sum
\begin{equation}
\sum_m Y_l^m(\theta,\phi)Y^{m*}_{l}(\theta,\phi) = {2l+1\over 4\pi}
\end{equation}
we obtain,
\begin{equation}
\langle f(\theta,\phi)\rangle
\approx 
\sum_{l} {2l+1\over 4\pi} C_l 
\left(1+2 A\hat\lambda\cdot\hat n\right)
\label{eq:dipole_mod1} 
\end{equation}
Hence if we expand the observed temperature square field, $f(\theta,\phi)$,
in spherical harmonics, 
\begin{equation}
 f(\theta,\phi) = \sum_{lm}A_{lm}Y_l^m(\theta,\phi).
\label{eq:T2harmonic}
\end{equation}
we should find a significant dipole contribution. 
The amplitude of the dipole would be related to $A$ with the 
direction equal to $\hat\lambda$. 
 This observable, therefore, provides a simple test 
of the dipole modulation model.

We also define another observable, $Q$, which sums temperature fluctuation
square
in a particular hemisphere. Let $N$ represent the total number of pixels in any
particular hemisphere of the real sky. We define, 
\begin{equation}
 Q = {1\over N}\sum_{i = 1}^{N} f(\theta,\phi)
\end{equation}
We divide the sky into an upper and lower hemisphere, along the chosen
$z$ direction. The resulting values of $Q$ in the two hemispheres are 
denoted as $Q_+$ and $Q_-$. We define the quantity, $S_Q$, such that,
\begin{equation}
 S_Q = {Q_+\over Q_-} 
\label{eq:SQ}
\end{equation}
This is maximized over all possible choices of the $z$ axis.  
The maximum value of $S_Q$ provides another useful statistic to test
for hemispherical anisotropy in real space. The corresponding $z$
axis gives an estimate of the dipole modulation axis $\hat \lambda$. 
We should point out that using this procedure we will also get contributions
 from higher order odd multipoles if they are present
in the harmonic expansion, 
Eq. \ref{eq:dipole_mod1}, of
 $f(\theta,\phi)$. Hence the results obtained by this procedure would,
in general, differ from the dipole extracted by making a harmonic
decomposition of $f(\theta,\phi)$. 

\subsection{Tests in multipole space}
 The two point correlations of the isotropic temperature field, $g(\hat n)$, 
 in two different directions, $ \hat n$ and $\hat n'$, are given as,
\begin{equation}
\left \langle{g(\hat n) g^*(\hat n')}\right \rangle = 
\sum_{lm}C_{l}Y_l^{m}(\hat n)Y_l^{m*}(\hat n')
\end{equation}
where $C_l$ is the angular power spectrum.  
The two point correlations  
of the dipole modulated temperature field, Eq. \ref{eq:dipole_mod}, 
can be expressed as,
\begin{equation}
\left \langle{\Delta T(\hat n) \Delta T^*(\hat n')} \right \rangle = 
\sum_{lm}C_lY_l^m(\hat n)Y_l^{m*}(\hat n')\left[1+A\cos{\theta} + A\cos\theta'
\right]
\label{eq:DTDT}
\end{equation}
where we drop terms quadratic in $A$. 
The two point correlation function of the 
spherical harmonic coefficients $a_{lm}$, defined in Eq. \ref{eq:DeltaT_exp}, 
can be expressed as,
\begin{equation}
\langle{a_{lm}a^*_{l'm'}}\rangle = \int d\Omega_{\hat n} d\Omega_{\hat n'}Y_l^{m*}(\hat n) Y_{l'}^{m'}(\hat n') \langle{{\bigtriangleup T}(\hat n)  
{\bigtriangleup T}(\hat n')}\rangle
\end{equation}
Using Eq. \ref{eq:DTDT}, we obtain,
\begin{equation}
\langle{a_{lm}a^*_{l'm'}}\rangle = \langle{a_{lm}a^*_{l'm'}}\rangle_{iso} +\langle{a_{lm}a^*_{l'm'}}\rangle_{dm} 
\end{equation}
where $\langle{a_{lm}a^*_{l'm'}}\rangle_{iso} = C_l\delta_{ll'}\delta_{mm'}$, and 
\begin{equation}
 \langle{a_{lm}a^*_{l'm'}}\rangle_{dm} = AC_{l'}\xi^{0}_{lm;l'm'}+AC_l\xi^{0}_{lm;l'm'}
\end{equation}
where,
\begin{eqnarray}
  \xi^{0}_{lm;l'm'} & = &\int d\Omega Y_l^{m*}(\hat n)Y_{l'}^{m'}(\hat n)\cos{\theta} \nonumber\\
&= &\delta_{m',m}\Bigg[\sqrt{\frac{(l-m+1)(l+m+1)}{{(2l+1)}{(2l+3)}}}\delta_{l',l+1}\nonumber\\
&&+\sqrt{\frac{(l-m)(l+m)}{{(2l+1)}{(2l-1)}}}\delta_{l',l-1}\Bigg]\ .
\end{eqnarray}
For $l' = l+1$,
\begin{equation}
 {\langle{a_{lm}a^*_{l'm'}}\rangle} = A(C_{l+1}+C_l)\delta_{m',m}\Bigg[\sqrt{\frac{(l-m+1)(l+m+1)}{{(2l+1)}{(2l+3)}}}\delta_{l',l+1}\Bigg]
\label{eq:corr1}
\end{equation}
and for $l' = l-1$,
\begin{equation}
 {\langle{a_{lm}a^*_{l'm'}}\rangle} = A(C_{l-1}+C_l)\delta_{m',m}\Bigg[\sqrt{\frac{(l-m)(l+m)}{{(2l+1)}{(2l-1)}}}\delta_{l',l-1}\Bigg]
\label{eq:corr2}
\end{equation}
Hence the dipole modulation model predicts a  
correlation between $a_{l,m}$
and $a_{l+1,m}$. 
We also find that the dipole modulation term does not give any contribution
to the power $\langle a_{lm}a^*_{lm}\rangle$ if we only retain terms
linear in $A$. 
Hence the power, $C_l$, of the field $\Delta T(\hat n)$ is same as that of 
$g(\hat n)$.  
The correlation in Eq. \ref{eq:corr2} may be positive or negative, depending
on the sign of $A$. However we expect that all the multipoles would
be correlated with the same sign. 
We can test for this correlation in real data by defining,
\begin{equation}
 C_{l,l+1} = \frac{l(l+1)}{2l+1}\sum_{m = -l}^{l} a_{lm}a^*_{l+1,m}
\label{eq:corr3}
\end{equation}
The sum over $m$ contains $(2l+1)$ terms. After dividing by $(2l+1)$
we obtain an average value of the correlation for each multipole. 
The normalization factor $l(l+1)$ is inserted in order to make 
it consistent with the measure $l(l+1)C_l$ used in  
\cite{Eriksen2004,Eriksen2007,Hansen2009,Erickce2008b,Paci2013,Schmidt2013}. 
We do not find a significant signal of anisotropy if the prefactor
$l(l+1)/(2l+1)$ is not used in the definition of $C_{l,l+1}$. 
The sum of $C_{l,l+1}$ over a chosen range of multipoles defines our 
statistic, $S_H(L)$,
\begin{equation}
S_H(L) = \sum_{l = 2}^{L} C_{l,l+1} 
\label{eq:SH}
\end{equation}

\subsubsection{Correlation Coefficient}
The correlation of the multipole coefficients, $a_{lm}$, using the dipole modulation temperature field is given by Eq. \ref{eq:corr1}.  
We can test this dependence by computing the Pearson correlation coefficient,
$r$, between $x$ and $y$, defined as,
\begin{eqnarray}
y &=& \sum_m a_{lm} a^*_{l+1,m}\nonumber\\
x &=&  \sum_m 
(C_{l+1}+C_l)\sqrt{\frac{(l-m+1)(l+m+1)}{{(2l+1)}{(2l+3)}}}
\label{eq:corr_xy}
\end{eqnarray}
These are defined by summing both sides in Eq. \ref{eq:corr1} over $m$ and
setting $l'=l+1$. 
Eq. \ref{eq:corr1} implies that,
\begin{equation}
y = Ax
\end{equation}
We compute the linear correlation coefficient to see whether there exists a 
linear correlation between these two variables over a certain range of
$l$ values. 
We again search over all directions in
order to maximize the correlation, $r$.

\section{Results}
In this section we present results for dipole modulation using 
the tests in real and multipole
space proposed above. 
 The significance of anisotropy
 is determined by comparing the result for real data with
that corresponding to 4000 random samples. The quoted P-values represent
the probability that the dipole extracted from real data may arise as
a random fluctuation from a statistically isotropic sample. 
We use the WMAP 9 year ILC map as well as the PLANCK \cite{Ade2013} NILC, SMICA
and SEVEM maps for our analysis. 

In most of our analysis we use masked sky
in order to eliminate the region close to 
galactic plane, which has very strong foreground
contamination. 
We use the $KQ85$ 
mask for the WMAP-ILC map and the COM-MASK-gal-07 mask for the PLANCK maps.
The masked regions are filled by randomly generated isotropic data. 
The random data is obtained by generating a full sky, random, isotropic,  
 high resolution map with $N_{side}=512$. 
After applying
the inverse mask to this map, it is added to the
masked real map at the same resolution. 
The resulting map is downgraded to $N_{side}=32$. 
This corresponds
to maximum $l$ value of 64, for which significant signal of hemispherical
anisotropy has been observed 
\cite{planck2013}.
The mask boundaries get smoothed when we downgrade the map and hence this
procedure eliminates any breaks that might have arisen if we worked
directly with the low resolution map.
We also test the sensitivity
of our results to mask boundaries by using an alternative procedure.
The PLANCK temperature and
the mask maps are provided at $N_{side} = 2048$. We generate the full sky
map at this resolution. 
This map is downgraded to $N_{side}=256$ after smoothing with a Gaussian
 beam having FWHM equal to
 three times of the pixel size of the low resolution map.
This map is then downgraded to $N_{side} = 32$.
This procedure gave results very close to the one described earlier.

 The results 
are found to depend significantly on the random realization used 
to fill the masked regions. 
Hence, the results for dipole power
and direction are obtained after averaging the results of 100 
maps, which are generated by filling the masked regions by
different random samples. 

\subsection{Dipole amplitude in real space}
We first consider the dipole extracted by making the spherical harmonic
decomposition, Eq. \ref{eq:T2harmonic}, of the temperature square field.  
If the field is statistically isotropic, none of the multipoles would
be significantly different from those corresponding to a random 
realization. If the dipole modulation is present, then we should detect
a significant dipole. Our calculations show that
 a direct extraction of the dipole
power from data does not yield a statistically significant result. 
The problem is traced to the fact that the temperature square dipole
power does not yield a sensitive 
probe of the dipole modulation model. 
 We check the sensitivity of this measure by direct simulations.
We generate a random isotropic realization of the CMB temperature map and 
multiply it with the dipole modulation factor, setting $A=0.072$. 
This generates a particular
realization of an anisotropic map. The significance of the measure
is determined by comparison with isotropic random samples, as explained above.
We do not find a statistically significant signal for $A=0.072$.    

\begin{figure}[!t]
\centering
\includegraphics[scale=0.80,angle=0]{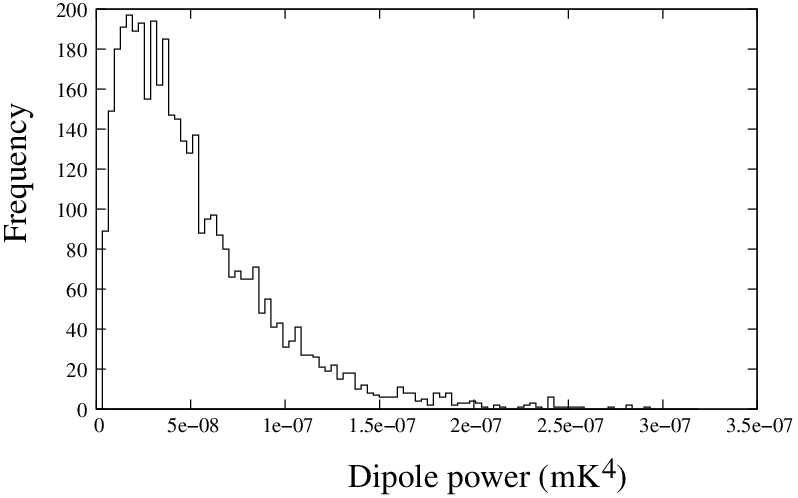}
\caption{The distribution of the dipole power, ${\cal C}_1(8)\ ({\rm mK}^4)$, 
of the temperature square
field for isotropic random realizations using cut (b). 
}
\label{fig:dipole_C1}
\end{figure}

The reason for this lack of sensitivity is easily understood.
The statistical measure used 
in \cite{Eriksen2004,Eriksen2007,Hansen2009,Erickce2008b,Paci2013,Schmidt2013} 
is the sum of $l(l+1) C_l$ over $l$. 
However in our case the dipole amplitude of the temperature
square field, Eq. \ref{eq:dipole_mod1}, essentially sums over $(2l+1)C_l$
and hence puts lower weight on higher multipoles. We, therefore, define
an alternate measure, by removing some of the low $l$ multipoles 
of the temperature field. We consider
two cuts: 
\begin{itemize}
\item[(a)] remove $l=2-4$ multipoles 
from the temperature field. 
\item[(b)] remove $l=2-8$ multipoles 
from the temperature field. 
\end{itemize}
The resulting dipole power of the temperature
square field, $f(\theta,\phi)$, for the two cuts (a) and (b) is denoted as, 
${\cal C}_1(4)$ and ${\cal C}_1(8)$ respectively.
The dipole power 
and the corresponding
direction parameters, $(\theta,\phi)$, are given in Tables \ref{tb:dipole1}
and \ref{tb:dipole3} for ${\cal C}_1(4)$ and ${\cal C}_1(8)$ respectively. 
Here we extract these parameters using the full sky map, i.e. without
masking the galactic plane. 
The significance of the dipole is given in terms of P-values, defined
above. 
In Fig. \ref{fig:dipole_C1}, we show the distribution of the
 dipole power, ${\cal C}_1(8)$, for isotropic random samples.

\begin{figure}[!t]
\centering
\includegraphics[scale=0.40,angle=270]{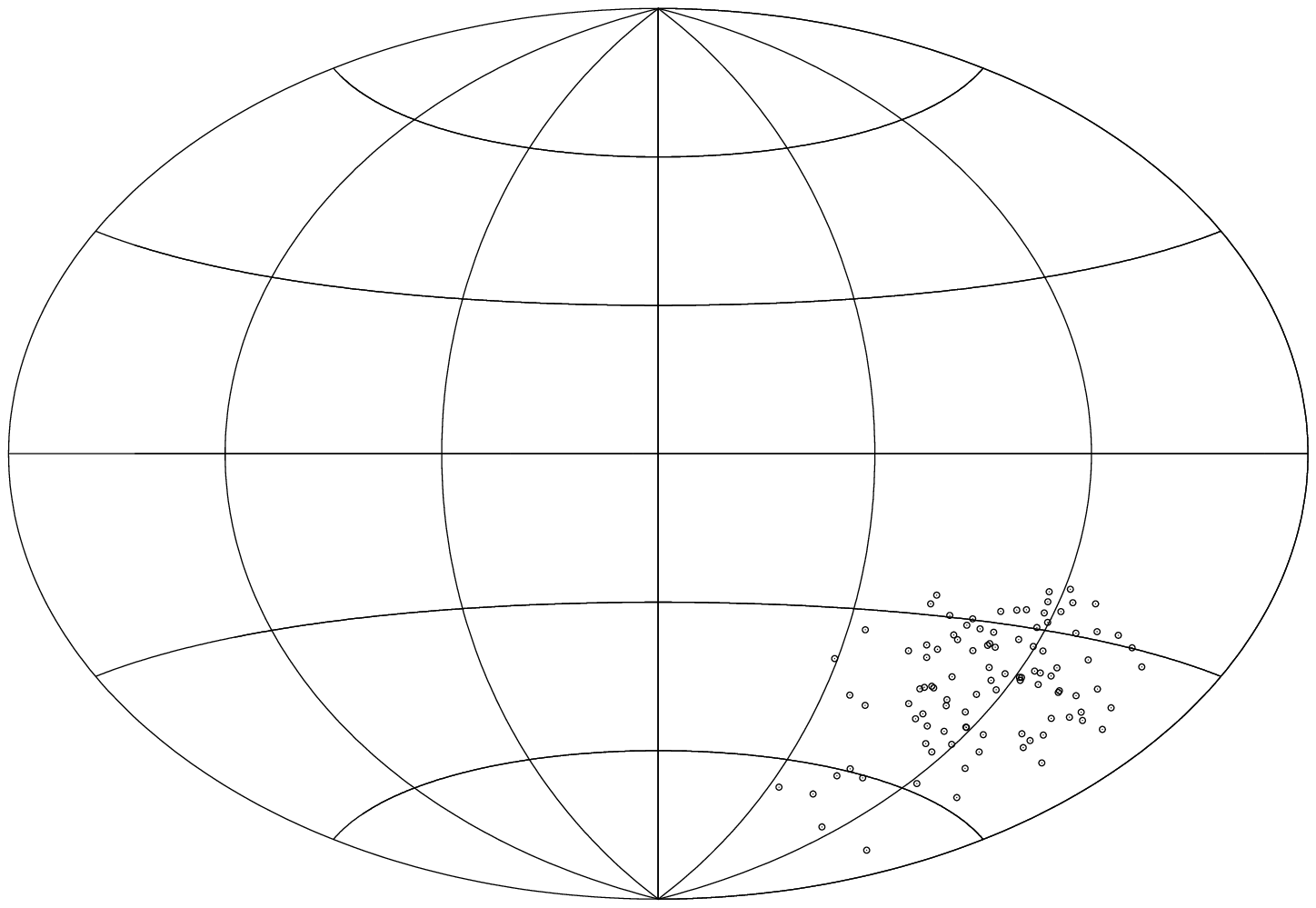}
\caption{The dipole axes parameters extracted from the PLANCK 
SMICA map using 
cut (b). The different 
 axes correspond to different random fillings of the masked regions. 
 }
\label{fig:axes}
\end{figure}

We next present results obtained by using masked sky analysis. 
 In Tables \ref{tb:dipole2} and \ref{tb:dipole4}, 
we present the results for ${\cal C}_1(4)$, cut (a), and ${\cal C}_1(8)$,
cut (b), respectively, along with 
the corresponding direction parameters and P-values. We find a signal 
of anisotropy with significance ranging from $2-3$ sigmas.  
As explained above, the results given in
Tables \ref{tb:dipole2} and \ref{tb:dipole4} are obtained after 
 taking a mean over 100 different realizations.
The variation of the axes for these 100 realizations for SMICA using cut
(b) is shown in Fig. \ref{fig:axes}. The mean values of $(\theta,\phi)$
in this case are found to be $(130^\circ, 244^\circ)$ and the corresponding
standard deviations $(11^\circ, 23^\circ)$. Hence we find that our extracted
axis parameters are consistent with those found in 
\cite{Eriksen2004,Eriksen2007,Hansen2009,Erickce2008b,Paci2013,Schmidt2013}. 
Similar spread in axes parameters are found with other maps.

\begin{table}[h!]
\begin{tabular}{|c|c|c|c|}
\hline
 &  ${\cal C}_1(4)\ ({\rm mK}^4)$ & $(\theta,\phi)$ & P-value\tabularnewline
\hline
NILC &  $5.11\times 10^{-7}$ & $(130^\circ, 230^\circ)$ & $2.92$\%\tabularnewline
\hline
SMICA & $6.96\times 10^{-7} $ & $ (122^\circ, 215^\circ)$ & $0.82$\%\tabularnewline
\hline
SEVEM  &$4.17 \times 10^{-7} $ & $(132^\circ, 237^\circ)$ &  $5.82\%$\tabularnewline
\hline
WMAP-ILC &$7.42 \times 10^{-7} $ & $(123^\circ,213^\circ)$ &  $0.57\%$\tabularnewline
\hline
\end{tabular}
\caption{The extracted dipole power of the temperature square field and the 
corresponding direction parameters, for different maps, using $N_{side}=32$, 
without masking the galactic plane. Here we have imposed cut (a) on the 
 temperature field.}
\label{tb:dipole1}
\end{table}

\begin{table}[h!]
\begin{tabular}{|c|c|c|c|}
\hline
 &  ${\cal C}_1(8)\ ({\rm mK}^4)$ & $(\theta,\phi)$ & P-value\tabularnewline
\hline
NILC &  $1.43\times 10^{-7}$ & $(124^\circ, 287^\circ)$ & $4.05$\%\tabularnewline
\hline
SMICA & $1.42\times 10^{-7} $ & $ (125^\circ, 236^\circ)$ & $4.25$\%\tabularnewline
\hline
SEVEM  & $1.51 \times 10^{-7} $ & $(127^\circ, 285^\circ)$ &  $3.45\%$\tabularnewline
\hline
WMAP-ILC &$1.21 \times 10^{-7} $ & $(131^\circ,241^\circ)$ &  $5.77\%$\tabularnewline
\hline
\end{tabular}
\caption{The extracted dipole power of the temperature square field and the 
corresponding direction parameters for cut (b), using $N_{side}=32$, 
without masking the galactic plane.}
\label{tb:dipole3}
\end{table}

\begin{table}[h!]
\begin{tabular}{|c|c|c|c|}
\hline
 &  ${\cal C}_1(4)\ ({\rm mK}^4)$ & $(\theta,\phi)$ & P-value\tabularnewline
\hline
NILC &  $4.07\times 10^{-7}$ ($2.91\times 10^{-7}$) & $(135^\circ,226^\circ)$ & $6.05\%$\tabularnewline
\hline
SMICA & $4.15\times 10^{-7} $ ($3.01\times 10^{-7}$)& $(135^\circ,224^\circ)$ & $5.85\%$\tabularnewline
\hline
SEVEM  &$3.93 \times 10^{-7} $ ($2.80\times 10^{-7}$)& $(136^\circ,229^\circ)$ & $6.82\%$\tabularnewline
\hline
WMAP-ILC &$4.06 \times 10^{-7} $ ($2.28\times 10^{-7}$) & $(131^\circ,216^\circ)$ & $6.65\%$\tabularnewline
\hline
\end{tabular}
\caption{The extracted dipole power of the temperature square field and 
the direction parameters for different maps, using masked sky analysis
with $N_{side}=32$, for cut (a).
The ${\cal C}_1$ values are obtained after taking a mean of 100
different realizations, with the masked regions filled by random isotropic
data. The corresponding standard deviations are given in brackets.}
\label{tb:dipole2}
\end{table}

We also simulate the dipole modulated field by taking the dipole amplitude $A = 0.072\pm 0.022$ and
extract the power, ${\cal C}_1$. For the case of masked sky, we find that the 
dipole amplitude, ${\cal C}_1(4) = (3.3\pm 1.1)\times 10^{-7} \ ({\rm mK}^4)$
and ${\cal C}_1(8) = (1.4\pm 0.7)\times 10^{-7} \ ({\rm mK}^4)$. 
Here the error arises due to the error in the extracted dipole amplitude, 
$|\Delta A|=0.022$. Hence we find that the theoretical expectations agree, within errors,
with the extracted dipole power from data.
Our results suggest that there exists a significant signal for the
dipole modulation model, Eq. \ref{eq:dipole_mod}, 
in the CMBR data.

\begin{table}[h!]
\begin{tabular}{|c|c|c|c|}
\hline
 &  ${\cal C}_1(8)\ ({\rm mK}^4)$ & $(\theta,\phi)$ & P-value\tabularnewline
\hline
NILC &  $2.04 \times 10^{-7}$ ($1.15\times 10^{-7}$) & $(130^\circ,245^\circ)$ & $0.92\%$\tabularnewline
\hline
SMICA & $2.06\times 10^{-7} $ ($1.16\times 10^{-7}$)& $(130^\circ,244^\circ)$ & $0.89\%$\tabularnewline
\hline
SEVEM  &$2.05 \times 10^{-7} $ ($1.14\times 10^{-7}$)& $(130^\circ,246^\circ)$ & $0.92\%$\tabularnewline
\hline
WMAP-ILC &$1.8 \times 10^{-7} $ ($1.15\times 10^{-7}$) & $(126^\circ,242^\circ)$ & $1.45\%$\tabularnewline
\hline
\end{tabular}
\caption{The extracted dipole power of the temperature square field and 
the direction parameters for cut (b), using $N_{side}=32$, after masking the galactic plane.
The ${\cal C}_1$ values are obtained after taking a mean of 100
different realizations, with the masked regions filled by random isotropic
data. The corresponding standard deviations are given in brackets.
}
\label{tb:dipole4}
\end{table}

\subsection{Hemispherical power anisotropy in real space}
We next apply our hemispherical power anisotropy test in real space. 
In this case we compute the power  
by squaring the temperature fluctuation field in each hemisphere 
in pixel space.
We generate mask files for the upper hemisphere ($0<\theta\leq90^0$) and 
the lower hemisphere ($\theta >90^0$). 
The resulting power obtained after applying these masks is normalized with the 
number of pixels in the respective hemispheres.
The ratio of powers in the two hemispheres gives us an estimate of 
$S_Q$, defined in Eq. \ref{eq:SQ}. 
In this case we present results only after eliminating the contaminated
galactic plane by using the masks $KQ85$ and COM-MASK-gal-07 for WMAP and
PLANCK data respectively.
The statistic $S_Q$ is determined
directly from the masked sky. 

With this measure also, we do not obtain significant results unless we impose
cuts to eliminate a few low multipoles of the temperature field. 
The results after imposing  cuts (a) and (b)
 are given in Tables \ref{tb:hemi3} and \ref{tb:hemi4} respectively.
Here the angle parameters $(\theta,\phi)$ are determined by searching
over all possible directions in order to maximize $S_Q$. 
In Tables \ref{tb:hemi3} and \ref{tb:hemi4},
the P-values in brackets are obtained by determining the number of times
an isotropic  random sample yields an $S_Q$ larger than real data.
We use 4000 random samples for this purpose.
We do not explicitly search over directions for each random sample. 
Hence the significance obtained by this procedure would 
provide an overestimate. In order to account
for the search, we assume Gaussian statistics, and determine the P-value
corresponding to two additional search parameters. This yields the 
true P-values given in Tables \ref{tb:hemi3} and \ref{tb:hemi4}. These indicate that the
significance of dipole anisotropy in this case is about 2 sigma for PLANCK data,
a marginally significant result. The WMAP-ILC map yields a much higher
significance for cut (a), roughly equal to 3.7 sigmas. 
We also 
find that the direction obtained aligns reasonably well with that corresponding
to hemispherical power asymmetry in multipole space 
\cite{Eriksen2004,Eriksen2007,Hansen2009,Erickce2008b,Paci2013,Schmidt2013}.

The dipole modulation model with $A=0.072\pm 0.022$ leads to 
a maximum $S_Q$ value of $1.24\pm 0.08 $, both for cut (a) and (b). Hence
the value extracted from PLANCK data for cut (b) agrees with this prediction, within errors.
The value corresponding to cut (a) is found to be slightly larger than 
predicted. The WMAP data, however, gives somewhat larger result. In particular
the WMAP result for cut (a) deviates from the predicted value by more than
3 sigmas.

\begin{table}[h!]
\begin{tabular}{|c|c|c|c|}
\hline
 &  max. $S_Q$& $(\theta,\phi)$  & P-value\tabularnewline
\hline
NILC & 1.36 & $(94^\circ,236^\circ)$  & 5.4\% (0.57\%) \tabularnewline
\hline
SMICA & 1.35 & $(94^\circ,236^\circ)$  & 6.4\% (0.7\%) \tabularnewline
\hline
SEVEM  & 1.34 & $(94^\circ,236^\circ)$ & 6.4\% (0.7\%) \tabularnewline
\hline
WMAP-ILC &1.53 & $(135^\circ,202^\circ)$ & 0.022\% (0.001\%) \tabularnewline
\hline
\end{tabular}
\caption{The extracted values corresponding to the hemispherical
power asymmetry with cut (a), after masking the galactic plane, 
 for different maps using $N_{side}=32$. The P-values in brackets 
do not account for the search over the axes parameters.}
\label{tb:hemi3}
\end{table}

\begin{table}[h!]
\begin{tabular}{|c|c|c|c|}
\hline
 &  max. $S_Q$& $(\theta,\phi)$  & P-value\tabularnewline
\hline
NILC & 1.27 & $(135^\circ,202^\circ)$  & 5.2\% (0.55\%) \tabularnewline
\hline
SMICA & 1.27 & $(135^\circ,202^\circ)$  & 4.9\% (0.50\%) \tabularnewline
\hline
SEVEM  & 1.27 & $(102^\circ,232^\circ)$ & 5.0\% (0.52\%) \tabularnewline
\hline
WMAP-ILC & 1.39 & $(135^\circ,202^\circ)$ & 4.9\% (0.05\%) \tabularnewline
\hline
\end{tabular}
\caption{The extracted values corresponding to the hemispherical
power asymmetry for cut (b), after masking the galactic plane,
for different maps using $N_{side}=32$.}
\label{tb:hemi4}
\end{table}

\subsection{Correlations in multipole space}
The dipole modulation model leads to correlations among different
multipoles, given in Eq. \ref{eq:corr1} or \ref{eq:corr2}. 
We can, therefore, test the model by computing the significance of
these correlations in data.
We compute the correlation, $C_{l,l+1}$, defined in Eq. \ref{eq:corr3}, 
at a particular $l$ value. The statistic is then obtained by
summing over a certain range of $l$ values as in Eq. \ref{eq:SH}. Following the range of
observed hemispherical power asymmetry \cite{Hoftuft2009,planck2013}, we consider the 
  multipole range $l = 2-64$.
We extract the maximum value of $S_H(L)$ for different maps after
searching over all possible directions. Note that for the multipole range
$l=2-64$, $L=63$. The true P-values are extracted,
as in section 4.2, after accounting for the search over the angle
parameters. The results, after masking the galactic plane, are given in 
Table \ref{tb:corr_l2}.
In this case also the masked regions are filled by random isotropic samples. 
The results are obtained after averaging over 100 such realizations. 
The mean and standard deviations of the extracted $S_H(L)$ are given
in Table \ref{tb:corr_l2}. For SMICA, the standard deviations of the
angle parameters, $(\theta,\phi)$, are $(21^\circ,34^\circ)$. 
We find a signal of anisotropy with significance approximately 
equal to 2.4 sigmas.
The dipole modulation model, with 
 $A=0.072\pm 0.022$, yields a value of $S_H(L)$ equal to 
$0.023\pm 0.008$. Hence in this case also the value extracted from data
agrees well with this prediction.

\begin{table}[h!]
\begin{tabular}{|c|c|c|c|}
\hline
 &  max. $S_H(L)$ $({\rm mK}^2)$ & $(\theta,\phi)$ & P-value\tabularnewline
\hline
NILC & $2.25\times10^{-2}\ (0.53\times 10^{-2})$ & $(115^\circ, 234^\circ)$ & 1.5\% ($0.12\%$) \tabularnewline
\hline
SMICA & $2.26\times10^{-2}\ (0.54\times 10^{-2})$ & $(115^\circ, 232^\circ)$ & 1.5\% ($0.12\%$) \tabularnewline
\hline
SEVEM  & $2.24\times10^{-2}\ (0.53\times 10^{-2})$ & $(114^\circ, 234^\circ)$ & 1.5\% ($0.12\%$) \tabularnewline
\hline
WMAP-ILC &$2.22\times10^{-2}\ (0.66\times 10^{-2})$ & $(104^\circ, 227^\circ)$ & 1.3\% ($0.10\%$)  \tabularnewline
\hline
\end{tabular}
\caption{The extracted maximum values of $S_H(L)$ for different maps using $N_{side}=32$, using
masked sky analysis over the multipole range $l=2-64$. 
The standard deviations of $S_H(L)$ are given in brackets.
The P-values given in brackets do not account for the search
over the two axes parameters.
}
\label{tb:corr_l2}
\end{table}

{\it Pearson correlation coefficient:}
We finally compute the Pearson correlation coefficient, $r$, between variables
$x$ and $y$, defined in 
section 3.1, over the range $l=2-64$, 
in order to test the relationship, Eq. \ref{eq:corr1}.
Here again we make a search over all directions in order to maximize 
the value of $r$. 
The results for cuts (a) and (b) are given in Tables 
\ref{tb:corr_r3} and
\ref{tb:corr_r4} respectively. In this case, cut (a) does not yield a
significant results whereas cut (b) yields a marginally significant signal. 
Furthermore, the direction parameters also show larger deviation 
in comparison to those found in hemispherical multipole space analysis
\cite{Eriksen2004,Eriksen2007,Hansen2009,Erickce2008b,Paci2013,Schmidt2013}. 
For SMICA, the standard deviations of $(\theta,\phi)$ are found to be
$(23^\circ,74^\circ)$, which are quite large in comparison to remaining tests. 
The dipole modulation model, 
Eq. \ref{eq:dipole_mod},
 with $A=0.072\pm0.022$ yields the maximum value of $r$ equal to
approximately 0.2. Hence our analysis gives a much larger value in
comparison to this prediction. This suggests that this measure may
not be a sensitive probe and gets a large contribution due to fluctuations
in data. Hence the discrepancy found by this probe is not very significant.

\begin{table}[h!]
\begin{tabular}{|c|c|c|c|}
\hline
 &  max. $r$ & $(\theta,\phi)$ & P-value\tabularnewline
\hline
NILC & 0.61 (0.11)  & $(163^\circ, 219^\circ)$ & 26\% ($4.4\%$) \tabularnewline
\hline
SMICA & 0.61 (0.11)  & $(162^\circ, 215^\circ)$ & 25\% ($4.3\%$) \tabularnewline
\hline
SEVEM  & 0.61 (0.11) & $(164^\circ, 229^\circ)$ & 27\% ($4.7\%$) \tabularnewline
\hline
WMAP-ILC & 0.66 (0.09)& $(150^\circ, 208^\circ)$ & 15\% ($2.2\%$)  \tabularnewline
\hline
\end{tabular}
\caption{The extracted maximum values of the 
correlation coefficient for different maps with cut (a)
using $N_{side}=32$, over the range $l=2-64$,
using masked sky analysis.
The standard deviation of $r$ is given in brackets.
}
\label{tb:corr_r3}
\end{table}

\begin{table}[h!]
\begin{tabular}{|c|c|c|c|}
\hline
 &  max. $r$ & $(\theta,\phi)$ & P-value\tabularnewline
\hline
NILC & 0.57 (0.10)  & $(130^\circ, 261^\circ)$ & 6.9\% ($0.77\%$) \tabularnewline
\hline
SMICA & 0.57 (0.10) & $(130^\circ, 261^\circ)$ & 6.9\% ($0.77\%$) \tabularnewline
\hline
SEVEM  & 0.57 (0.10)  & $(130^\circ, 263^\circ)$ & 6.7\% ($0.75\%$) \tabularnewline
\hline
WMAP-ILC & 0.53 (0.09) & $(134^\circ, 260^\circ)$ & 12\% ($1.52\%$)  \tabularnewline
\hline
\end{tabular}
\caption{The extracted maximum values of the
correlation coefficient for different maps using cut (b) 
and $N_{side}=32$, over the range $l=2-64$,
using masked sky analysis.
The standard deviation of $r$ is given in brackets.}
\label{tb:corr_r4}
\end{table}

\section{Dependence of dipole amplitude on $l$}
\begin{figure}[!t]
\centering
\includegraphics[scale=0.60,angle=270]{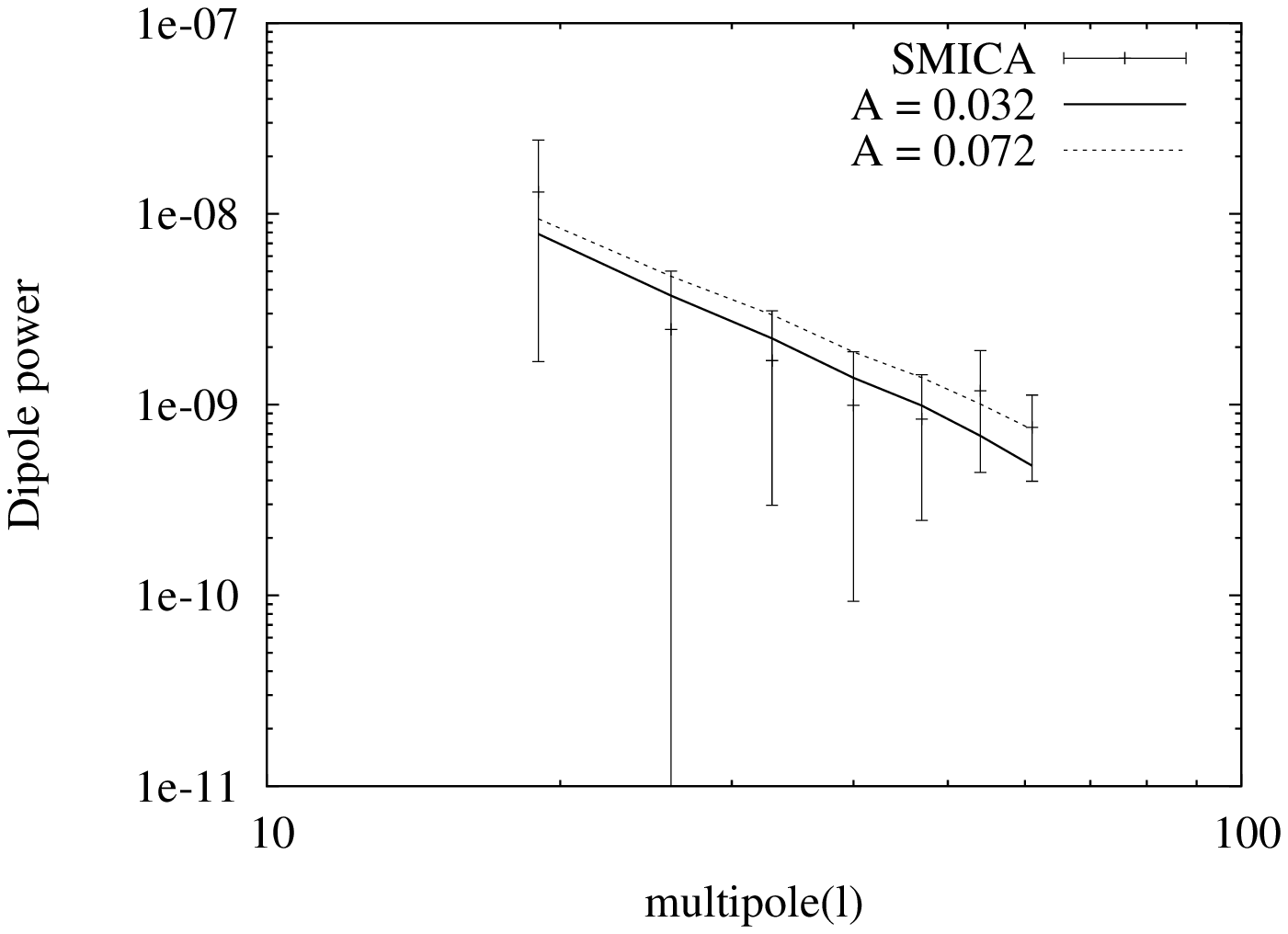}
\caption{The dipole power, ${\cal C}_1\ ({\rm mK}^4)$, 
of the temperature square
field as a function of the multipole $(l)$. The solid and dotted straight
lines show the theoretical prediction based on $A=0.032$ and $A=0.072$
respectively. 
}
\label{fig:dipole_C1A}
\end{figure}

We next determine how the dipole modulation amplitude $A$ varies with
the multipole $l$. For this analysis we extract the effective value of
$A$ over a narrow range of $l$ values. These are taken to be,
$l=2-8,9-15, ..., 58-64$. The extracted dipole power of the temperature
square field, using the SMICA map, is shown in Fig. \ref{fig:dipole_C1A}. 
For comparison we also show results for simulated maps corresponding
to $A=0.072$ and $A=0.032$. Here we do not show the results for $l=2-8, 9-15$
since the fluctuations in this multipole range are found to be very large. 
Hence it is not possible to make a reliable comparison of theoretical
prediction with data. We see from Fig. \ref{fig:dipole_C1A} that the
standard value $A=0.072$ gives a good description of the high $l$ multipoles.
However the low $l$ multipoles prefer a value closer to $A=0.032$. Hence
we find that the effective value of $A$ increases with $l$.  
A more detailed investigation of this effect is postponed for future research.

\section{Discussion and Conclusions}
We find a significant signal of the dipole modulation model, 
Eq. \ref{eq:dipole_mod}, in data. 
In real space we propose a measure based on the dipole power
of the temperature square field. 
This measure reveals a significant signal of anisotropy
provided we eliminate a few low $l$ multipoles. The extracted
signal in this case essentially sums the $C_l$ values over $l$ after
weighting them with $(2l+1)$. In comparison to the standard measure, 
$l(l+1) C_l$, used in 
\cite{Eriksen2004,Eriksen2007,Hansen2009,Erickce2008b,Paci2013,Schmidt2013},
this yields a lower weight for high $l$ multipoles. 
By eliminating a few low $l$ multipoles, we find that our measure also
provides a sensitive probe of the dipole modulation model. The extracted
values, both the dipole amplitude and direction are found to be in 
agreement, within errors,
 with the prediction based on multipole space hemispherical
analysis. In multipole space, we show that the dipole modulation model
leads to 
correlations between multipoles which differ by $\Delta l=1$.  
We define a suitable measure of this correlation, which also reveals a  
statistically significant signal of anisotropy. The amplitude and
 direction is again found to be in agreement with expectations.
In multipole space, we also predict a linear dependence between the $x$ and $y$
variables defined in Eq. \ref{eq:corr_xy}. This test is not found to be
a sensitive probe of the signal, even after eliminating low $l$
multipoles. Nevertheless, it leads to a marginally significant result after
eliminating a few low $l$ multipoles. 
We perform a preliminary analysis in order to test the dependence of
the dipole modulation amplitude $A$ on $l$. We find that the effective 
value of $A$ increases with $l$ in the range $l=2-64$. 
Our results suggest that the hemispherical anisotropy found in 
\cite{Eriksen2004,Eriksen2007,Hansen2009,Erickce2008b,Paci2013,Schmidt2013} 
can be consistently attributed to the 
dipole modulation model, Eq. \ref{eq:dipole_mod}. 

{\bf Acknowledgments:}
We acknowledge the use of WMAP data available from NASA's LAMBDA site(http://lambda.gsfc.nasa.gov/).
Some of the results in this paper have been derived using the HEALPix
\cite{Gorski2005} package.

\end{document}